\newif\ifanonymous
\newif\iftodo

\documentclass[runningheads]{llncs}

\usepackage[nolist,printonlyused]{acronym}
\usepackage{hyperref}
\usepackage{listings}
\usepackage{booktabs}
\usepackage{xurl}
\usepackage{orcidlink}

\usepackage[%
	group-minimum-digits=3,
	round-mode=places,
	round-precision=2,
	round-pad=false]{siunitx}

\usepackage{multirow}
\usepackage{xurl}
\usepackage{graphicx}
\usepackage{dirtytalk}
\usepackage[
	textsize=small,
	backgroundcolor=red!50!white,
	linecolor=red!50!white]{todonotes}
\usepackage{amsmath}
\usepackage{amssymb}
\usepackage{colortbl}

\colorlet{positive}{green!60!white}
\colorlet{negative}{red!40!white}
\colorlet{fg}{positive}
\colorlet{bg}{negative}

\usepackage[capitalise]{cleveref}
\crefname{definition}{Def.}{Def.}
\crefformat{section}{\S#2#1#3}

\newcommand{\pull}{\ensuremath{\mathit{pull}}}

\def\eg{\emph{e.g.}}
\def\ie{\emph{i.e.}}
\def\cf{\emph{cf.}}

\def\see#1{(\cf~\cref{#1})}

\def\rq#1{\textbf{(RQ#1)}}
\def\c#1{\textbf{(C#1)}}

\def\makeanon#1{\ifanonymous[ANONYM]\else#1\fi}

\def\threatcrawl{\textsc{ThreatCrawl}}

\def\tcbfk{\texttt{TC\_BFK}}
\def\tcbf{\texttt{TC\_BF}}
\def\tcfk{\texttt{TC\_FK}}
\def\tcbk{\texttt{TC\_BK}}
\def\tcf{\texttt{TC\_F}}
\def\tcb{\texttt{TC\_B}}
\def\tck{\texttt{TC\_K}}

\def\pages{\ensuremath{\mathbb{P}}}
\def\crawl{\ensuremath{\mathit{crawl}}}

\newcommand{\abs}[1]{\ensuremath{\lvert #1 \rvert}}
\newcommand{\mymax}[1]{\ensuremath{\lceil #1 \rceil}}

\newcommand{\avg}[1]{\ensuremath{\overline{#1}}}

\def\hlp{\cellcolor{positive}}
\def\hln{\cellcolor{negative}}

\begin{acronym}
	\acro{apt}[APT]{Advanced Persistent Threat}
	\acro{bert}[BERT]{Bidirectional Encoder Representations from Transformers}
	\acro{bfc}[BFC]{Bootstrapping Focused Crawlers}
	\acro{bow}[BoW]{Bag of Words}
	\acro{cti}[CTI]{Cyber Threat Intelligence}
	\acro{cve}[CVE]{Common Vulnerabilities and Exposures}
	\acro{cert}[CERT]{Computer Emergency Response Team}
	\acro{dom}[DOM]{Document Object Model}
	\acro{dos}[DOS]{denial-of-service}
	\acro{html}[HTML]{HyperText Markup Language}
	\acro{ids}[IDS]{Intrusion Detection System}
	\acro{ioc}[IOC]{Indicator of Compromise}
	\acroplural{ioc}[IOCs]{Indicators of Compromise}
	\acro{ml}[ML]{Machine Learning}
	\acro{nlp}[NLP]{Natural Language Processing}
	\acro{oov}[OOV]{Out-of-Vocabulary}
	\acro{osint}[OSINT]{Open Source Intelligence}
	\acro{svm}[SVM]{Support Vector Machine}
	\acro{ttp}[TTP]{Tactics, Techniques \& Procedures}
	\acroplural{ttp}[TTPs]{Tactics, Techniques \& Procedures}
	\acro{tfidf}[TF-IDF]{Term Frequency-Inverse Document Frequency}
	\acro{url}[URL]{Uniform Resource Locator}
	\acro{pop}[PoP]{Pyramid of Pain}
	\acro{sbert}[SBERT]{Sentence-BERT}
	\acro{soc}[SOC]{Security Operations Center}
	\acro{cvss}[CVSS]{common vulnerability scoring system}
	\acro{mab}[MAB]{multi-armed bandit}
\end{acronym}

\begin{document}

\pagestyle{headings}

\title{Bandit on the Hunt: Dynamic Crawling for Cyber Threat Intelligence}
\titlerunning{Bandit on the Hunt}

\ifanonymous\else
	\author{
	Philipp Kuehn\inst{1}\orcidlink{0000-0002-1739-876X} \and
	Dilara Nadermahmoodi\inst{1} \and
	Markus Bayer\inst{1}\orcidlink{0000-0002-2040-5609} \and
	Christian Reuter\inst{1}\orcidlink{0000-0003-1920-038X}
}

\institute{%
	Science and Technology for Peace and Security (PEASEC), \\
	Technical University of Darmstadt, Germany
}

\authorrunning{P. Kuehn et al.}

\fi

\maketitle

\begin{abstract}
	Public information contains valuable \ac{cti} that is used to prevent future attacks.
	While standards exist for sharing this information, much appears in non-standardized news articles or blogs.
	Monitoring online sources for threats is time-consuming and source selection is uncertain.
	Current research focuses on extracting \aclp{ioc} from known sources, rarely addressing new source identification.
	This paper proposes a \ac{cti}-focused crawler using \ac{mab} and various crawling strategies.
	It employs \acs{sbert} to identify relevant documents while dynamically adapting its crawling path.
	Our system \threatcrawl{} achieves a harvest rate exceeding 25\% and expands its seed by over $300\%$ while maintaining topical focus.
	Additionally, the crawler identifies previously unknown but highly relevant overview pages, datasets, and domains.
\end{abstract}

\acresetall

\section{Introduction}
\label{sec:introduction}

In \ac{cti} information is used to learn from current threats and prevent similar attacks against infrastructures.
This is done by sharing actionable information such as \acp{ioc} through various channels.
It borrows its intelligence generation procedure from the intelligence cycle used by intelligence services~\cite{sauerwein_threat_2021}.
\ac{cti} has established standards to publish, import, and export information to databases and platforms~\cite{preuveneers_privacypreserving_2022}.
But \ac{cti} is often shared in unstructured formats like blog posts or threat reports~\cite{husari_ttpdrill_2017}.
Manually scanning online posts for \acp{ioc} is time-consuming for personnel.
Hinchy~\cite{hinchy_voice_2022} surveyed 468 full-time security analysts, finding over half spend most time on manual tasks, believe automation is possible, and may change jobs without modern tools.
\makeanon{Kaufhold et al.~\cite{kaufhold_we_2024}} confirmed these results with participants stating they lack capacity to monitor all media and need more automation.

\ac{ioc} extractors and threat detection methods are being developed~\cite{liao_acing_2016,lesceller_sonar_2017}, along with vulnerability severity predictors~\ifanonymous\cite[ANONYM]{elbaz_fighting_2020}\else\cite{elbaz_fighting_2020,kuehn_ovana_2021,kuehn_common_2023}\fi.
However, this research assumes the right information sources.
The Internet is dynamic, with sites changing focus, ceasing publication, or emerging anew.
Despite a decade of research, \ac{cti} cycle methods still fail to support practitioners in basic information collection~\cite{sauerwein_threat_2021,oosthoek_cyber_2021}.
Manual source identification conflicts with the existing information overload facing security personnel.

\def\crawlermainrq{How can \ac{cti} related information be identified and crawled from the web}

Maintaining current \ac{cti}-relevant websites requires unavailable manual work.
This creates blind spots for active attack campaigns.
Time spent searching for \ac{cti} results in less effective staff~\makeanon{\cite{kaufhold_we_2024}}.
This reduces infrastructure security as less time is spent on actual protection.
A precise approach to finding relevant web pages quickly is necessary.
Focused crawlers \say{selectively seek out pages that are relevant to a pre-defined set of topics}~\cite{chakrabarti_focused_1999}.
This work focuses on identifying \ac{cti}-related information published online.
We aim to answer the research question \textit{\say{\crawlermainrq?}~\rq{}}

This work combines techniques for crawling, classifying, and ranking content in our \threatcrawl{} pipeline.
It gathers specific \ac{cti} domain information from the surface web.
Our proposal uses \ac{sbert} embeddings to decide which sources to follow~\c{1}.
Documents are classified by information type and ranked by domain suitability~\c{2}.

First, we provide an overview of related work in~\cref{crawl:sec:related_work}.
\cref{crawl:sec:method} presents the theoretical concept of \threatcrawl{}, followed by initialization description in \cref{crawl:sec:threatcrawl}, while \cref{crawl:sec:evaluation} evaluates the system.
\cref{crawl:sec:discussion} discusses the evaluation, limitations and future work, and \cref{crawl:sec:conclusion} concludes this work.

\section{Related Work}
\label{crawl:sec:related_work}

Traditional document embeddings like \ac{tfidf} are surpassed by context-aware, deep-learning embeddings~\cite{devlin_bert_2018,peters_deep_2018}.
Reimers and Gurevych~\cite{reimers_sentencebert_2019} build on \ac{bert} with \ac{sbert} for generating sentence and document embeddings.

Several focused crawlers use \ac{tfidf} as embedding method~\cite{wang_focused_2010,liu_focused_2023,pham_bootstrapping_2019}.
Zhang et al.~\cite{zhang_dsdd_2021} propose finding datasets lacking metadata using a \ac{mab} focused crawler~\cite{pham_bootstrapping_2019}.
Koloveas et al.~\cite{koloveas_intime_2021} propose an integrated \ac{ml}-based crawler for managing \ac{cti} information using ACHE and Gensim.
Sanagavarapu et al.~\cite{sanagavarapu_siren_2018} propose a cybersecurity-specific search engine.

While some \ac{cti} research focuses on Twitter/X for easy access~\ifanonymous\cite[ANONYM]{tundis_featuredriven_2022}\else\cite{tundis_featuredriven_2022,riebe_cysecalert_2021}\fi, others use known sites~\cite{liao_acing_2016}.
Both methods rely on known sources without expanding view.
Since Twitter/X's leadership change, crawling it became more difficult.
Dekel et al.~\cite{dekel_mabat_2023} use \ac{mab} to prioritize investigated attacks in \ac{cti} datasets.

Current work combining web crawling, classification, and ranking for \ac{cti} in a single pipeline is widely discussed.
Tawil and Alqaraleh~\cite{tawil_bert_2021} describe a crawler using \ac{sbert} embeddings for document labeling.
This approach separates crawling and classifying processes, crawling everything before classification.
Koloveas et al.~\cite{koloveas_intime_2021} present a different two-step approach, reducing focused crawling benefits with a harvest rate of 9.5~\%.
\acp{cert} and \acp{soc} personnel already monitor a small set of domains defining their infrastructure scope.
An open question remains developing a general one-step focused crawling approach combining crawling, classification, and ranking of \ac{cti}-relevant content based on known URLs.

\section{Methodology}
\label{crawl:sec:method}

The goal of the present work is the identification of new, previously unknown web pages, that are related to the user's input URLs.
We combine the different concepts proposed in \cite{pham_bootstrapping_2019}, integrate new technology, and tailor them to the requirement of security personnel.
This work builds on top of \cite{kuehn_threatcrawl_2023}.

\subsection{Notation}
\label{crawl:sec:notation}

We denote \pages{} as the set of all web pages.
Given two pages $p, p' \in \pages$, $p \to p'$ indicate, that $p$ links to $p'$ and $p \sim_\theta p'$ indicates that $p$ is contextually similar to $p'$ with regard to a relevance threshold $\theta$.
$theta$ is omitted if it is clear from context.
We extend $\sim$ to sets, \ie, $p \sim_\theta P \subseteq \pages \iff \exists p' \in P. p \sim_\theta p'$.
Similarly, $P, P' \subseteq \pages$, $P \sim\theta P' \iff \forall p \in P. p \sim\theta P'$.
The function $\crawl_a(p) \to \mathcal{P}(\pages)$ denotes a crawl of page $p \in \pages$ based on a crawl action $a$, which returns a set of pages.

\subsection{Problem Definition}
\label{crawl:subsec:problem}
One of the key aspects of this system is the relevance of information.
Recent studies show a shift of relevance in the \ac{cti} domain\footnote{Based on an annual survey conducted by the SANS Institute\footnote{\url{https://www.sans.org}} in which they survey security professionals from various organizations.} from very detailed information like \acp{ioc} to broader information like threat or malware reports~\see{tab:most_useful_cti_types}.
Therefore, we calculate the relevance of pages based on their similarity to the used seed $s \in S \subseteq \pages$, rather than using binary information like the existence of \ac{ioc} information~\cite{liao_acing_2016}.
Our approach might miss dataset pages that present \ac{ioc} information, \eg, through simple lists, but this is already covered by~\cite{pham_bootstrapping_2019}.
\cref{crawl:def:threatcrawl} defines the problem, we aim to solve.

\begin{definition}[\acs*{cti}-focused crawler]
	\label{crawl:def:threatcrawl}
	Given a set of seed pages $S \subseteq \pages$ defining the scope of ones \ac{cti} infrastructure security information, we want to identify $P \subseteq \pages$, such that $P \sim S$.
\end{definition}

\begin{table}[t]
	\centering\small
	\begin{tabular}{@{} c @{\hspace{8pt}} p{.25\linewidth} @{\hspace{8pt}} p{.3\linewidth} @{\hspace{8pt}} p{.3\linewidth} @{}}
		\toprule
		             & \textbf{2020}      & \textbf{2021}                                      & \textbf{2022}                                      \\ \midrule
		\textbf{\#1} & \acsp*{ioc}        & Textual information about targeted vulnerabilities & Textual information of used malware                \\ \addlinespace
		\textbf{\#2} & TTPs               & Textual information of used malware                & Textual information about targeted vulnerabilities \\ \addlinespace
		\textbf{\#3} & Adversary analysis & \acsp*{ioc}                                        & Broad information about attacker trends            \\
		\bottomrule
	\end{tabular}
	\caption{Top 3 most useful \acf*{cti} types according to the SANS CTI surveys from the years 2020 to 2022~\cite{robertmlee_cyber_2020,brown_sans_2021,brown_sans_2022}.}
	\label{tab:most_useful_cti_types}
\end{table}

We propose an architecture we call \threatcrawl{}.
It uses \ac{sbert} embeddings, and a one-step approach, \ie, classification during the crawling process to adjust the crawling direction on the fly.
It implements similar retrieval methods as \cite{pham_bootstrapping_2019} combined with a UCB1-\ac{mab}.
The retrieval methods are
(i) forward~(F),
(ii) backlink~(B), and
(iii) keyword search~(K), \ie, $\crawl_a$ with $a \in \left[F, B, K\right]$.
Given a page $p \in \pages$.
\emph{Forward search} crawls all links provided on $p$, \ie, $\crawl_F(p) = \left\{ p' \in \pages\ |\ p \to p' \right\}$.
\emph{Backlink search} crawls all pages, which link to $p$, \ie, $\crawl_B(p) = \left\{ p' \in \pages\ |\ p' \to p \right\}$.
\emph{Keyword search} searches for pages containing keywords of $p$, \ie, $\crawl_K(p) = \left\{ p' \in \pages\ |\ p'\ \text{contains keyword of}\ p \right\}$.
Those methods are proven to provide a broad coverage during focus crawling~\cite{pham_learning_2018,chakrabarti_surfing_1999,diligenti_focused_2000}, while still providing the possibility to dive deep in a source through forward searches.

\section{\threatcrawl}
\label{crawl:sec:threatcrawl}

This section provides a general overview of the used system to answer our research question~\see{crawl:subsec:overview}, followed by insights into the search actions~\see{crawl:subsec:actions}, classification method~\see{crawl:subsec:classifier}, and \ac{mab} selection~\see{crawl:subsec:mab}.

\subsection{Overview}
\label{crawl:subsec:overview}

\begin{figure*}
	\centering
	\includegraphics[width=.8\linewidth]{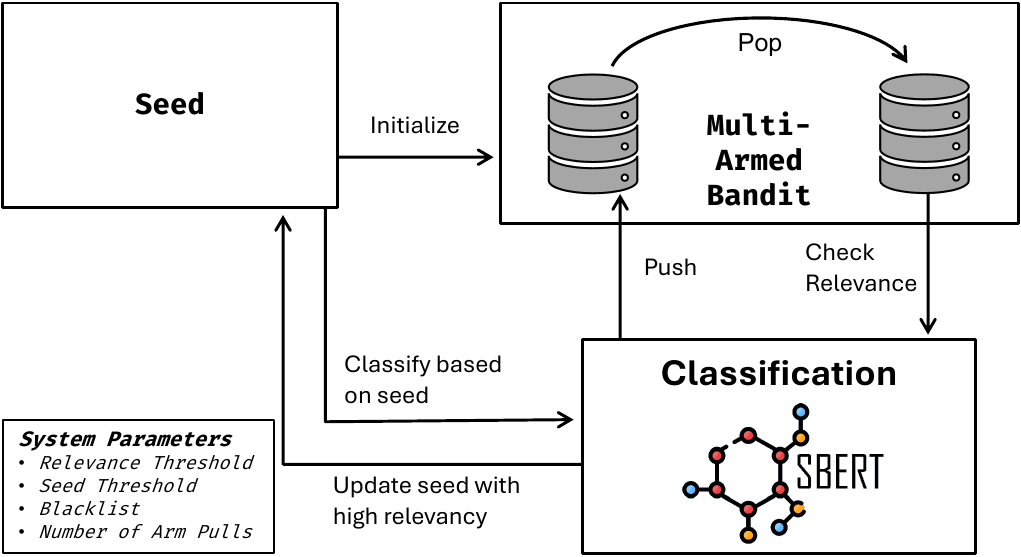}
	\caption{A schematic overview of our proposed \ac{cti} source identification system \threatcrawl{} leveraging a \acf*{mab} and classification based on \acf*{sbert}~\cite{reimers_sentencebert_2019}.}
	\label{crawl:fig:system}
\end{figure*}

We propose \threatcrawl{}~\see{crawl:def:threatcrawl} to answer our research question \emph{\say{\crawlermainrq{}}}.
Its central operation is bound on the user-provided set of seed URLs $S \subseteq \pages$ and the used crawling methods $F, B, K$.
An overview of the whole system is presented in \cref{crawl:fig:system}.
$S$ is crawled as ground truth for the classification and to provide the initial steps for crawling and prepare the \ac{mab}.
This is done using a priority queue based on the relevance of a given page.
For the seed pages $p \in S$ the priority is set to the highest possible value.
Following this step, the \ac{mab} is initialized with a discovery phase to calibrate the arm selection our search methods.
All pages that are discovered this way can be directly classified as relevant through their content and if relevant, pushed into the priority queue with their similarity to $S$ directly translated as priority, \ie, more similar pages are picked first.
Afterward, the priority queue is processed one step at a time, popping the most relevant page and using the \ac{mab} to select the most rewarding arm until one of the stop conditions is reached~\see{crawl:sec:method}.
This ensures the highest possible outcome measured as \emph{harvest rate}, \ie, $\frac{|R|}{|P|}$, where $R, P \subseteq \pages$ is the set of relevant pages and the set of all seen pages by the crawler, respectively.

\subsection{Search Actions}
\label{crawl:subsec:actions}

\paragraph{Forward Link Search}
Forward link search follows all hyperlinks on a page $p$, offering broad website coverage and uncovering in-depth content.

\paragraph{Backward Link Search}
Backward link search, commonly used in SEO tools, identifies pages linking back to $p$ to discover related content.
Since no open sources provide this data, we rely on a commercial API.

\paragraph{Keyword Search}
Keyword search extracts key terms from $p$ to guide searches, using tools like \emph{KeyBERT}~\cite{grootendorst_keybert_2021}, \emph{YAKE}, or \emph{RAKE}~\cite{amur_unlocking_2023}.
We use KeyBERT and combine the top three keywords with \texttt{OR} logic to ensure broader results.
The keyword search itself is done with commercial API.
% TODO: implement it using python's itertools.combinations with lenght n=2
% and query the search engine m times, where m is the number of resulting combinations (6 for 3 keywords)
% this would increase the coverage and stays more in focus with the keywords, as "vulnerability" or "proof-of-concept" are way to general

\subsection{Relevance Classification}
\label{crawl:subsec:classifier}

The content of a page $p$ is embedded using \ac{sbert}~\cite{reimers_sentencebert_2019}.
Pages and their embeddings are used interchangeably.
\Ac{sbert} provides dense vector representations capturing semantic similarity between pages.
Compared to \ac{tfidf} or Word2Vec, \ac{sbert} excels at contextual nuances and generates sentence-level embeddings better suited for semantic search tasks.
This allows contextual representation even where keyword-based approaches fail to capture deeper meanings.
Once embedded, cosine similarity between vectors of pages $p, q$ is calculated.
Cosine similarity measures vector similarity independent of magnitude, ideal for comparing web page semantics.
For pages $p, q$ and page set $Q$ we define $\mathit{sim}(p, q) = \text{cos\_sim}(p, q)$ and $\mathit{sim}(p, Q) = \max \left(\, [ \text{cos\_sim}(p, q) |\ q \in Q ]\, \right)$.
Two pages $p, q$ are semantically related $p \sim q \iff \mathit{sim}(p, q) \geq \theta$, for relevance threshold $\theta$.
A page $p$ and page set $Q$ are semantically related $p \sim Q \iff \mathit{sim}(p, Q) \geq \theta$.

\subsection{Multi-Armed Bandit}
\label{crawl:subsec:mab}

Multi-armed bandit algorithms are crucial for balancing exploration and exploitation in focused crawling.
We provide a comparison of \ac{mab} algorithms in \cref{crawl:tab:mab_comparison}.

\begin{table*}
	\caption{Comparison of \acl*{mab} algorithms.}
	\label{crawl:tab:mab_comparison}
	\centering
	\begin{tabular}{@{} p{.13\linewidth} @{\hspace{8pt}} p{.25\linewidth} @{\hspace{8pt}} p{.25\linewidth} @{\hspace{8pt}} p{.25\linewidth} @{}}
		\toprule
		\textbf{Algorithm} & \textbf{Best For}                                & \textbf{Strengths}                                        & \textbf{Weaknesses}                        \\
		\midrule
		Epsilon-Greedy     & Static or semi-static environments               & Simple to implement; tunable exploration                  & Inefficient in complex, large environments \\
		UCB1               & Stable environments, long-term performance       & Logarithmic regret, efficient exploration-exploitation    & Slow to adapt to dynamic changes           \\
		EXP3               & Adversarial or highly dynamic environments       & Robust to non-stationary rewards, no explicit exploration & Over-explores in well-structured settings  \\
		EXP3-IX            & Non-stationary but less adversarial environments & Better balance of exploration, robust                     & More complex to implement, requires tuning \\
		Sliding-Window UCB & Non-stationary, fast-changing environments       & Adapts to time-varying rewards                            & Needs careful window size tuning           \\
		\bottomrule
	\end{tabular}
\end{table*}

Based on \cref{crawl:tab:mab_comparison}, UCB1 is the best fit for our setting based on its strong performance in stable environments\footnote{Stable in terms of information it seeks, rather than the environment of the web as a whole, which is, as discussed, a dynamic environment.}, balancing exploration and exploitation with logarithmic regret.
It efficiently uses stable domain knowledge to optimize long-term performance without complex tuning.
Though slower to adapt to changes, its effectiveness in consistent environments makes it ideal for forward crawling tasks.
\cref{crawl:eq:reward} provides the reward function $r \in \mathbb{R}_{\geq 0}$ for a step $\pull(p)$ for page $p$, the set of seeds $S$, a domain weight $\delta$, and the number of discovered $|\mathit{domains}|$.

\begin{equation}
	\label{crawl:eq:reward}
	r(p) = \max (\delta\ |\mathit{domains}(\pull(p))|\ + \sum_{p'\, in\, \pull(p)} p' \sim S, 0)
\end{equation}

\section{Evaluation}
\label{crawl:sec:evaluation}

The evaluation was performed on a machine with an Intel Core i7-8565U processor and 16GB of RAM.
A relevance threshold of \num{0.6} was used to classify relevant pages.
The seed threshold was set to \num{0.8}, as a higher threshold of \num{0.9} resulted in nearly identical outcomes, which was undesirable for exploration.
A blacklist excluded social media platforms like X and YouTube, thread aggregators, non-HTTP protocols, and file types such as images, documents, and program-specific formats.
The number of steps was set to \num{500} and \num{2000} to allow sufficient exploration actions.
The seed consisted of \num{17} pages, covering security news and in-depth reports.

\subsection{Evaluation}

We evaluate our system based on the combinations of the different search actions, where
\textbf{B} indicates backward search,
\textbf{F} indicates forward search, and
\textbf{K} indicates keyword search.

\begin{description}
	\item[\tcbfk]
	      Balance broad coverage, deep relevance, and keyword-driven exploration.
	\item[\tcbf]
	      Follows forward links and analyzes backlinks to discover related pages, providing both broad and targeted exploration.
	\item[\tcfk]
	      Follows forward links and uses keyword searches to expand coverage, capturing more diverse content.
	\item[\tcbk]
	      Analyzes backlinks and performs keyword searches for targeted exploration and broader discovery.
	\item[\tcf]
	      Follows forward links from seed pages, offering extensive coverage but with potential inefficiencies.
	\item[\tcb]
	      Analyzes backlinks to uncover related pages not directly linked to the seed.
	\item[\tck]
	      Search solely based on keywords of given pages.
\end{description}

\begin{table*}
	\caption{
		Result of the evaluation with a maximum of \num{500} and \num{2000} steps.
		\abs{S}, \abs{P}, and \abs{P+} show the number of steps (if not the maximum), the number of total crawled pages, and number of relevant pages, of which are \abs{Seed} new seeds, HR the harvest rate [in \%], and \mymax{\sim} the maximum similarity of identified pages.
		\abs{Dom} and \abs{Dom+} indicate how many domains and relevant domains are identified, respectively.
		TM shows the top method during this run with \avg{\text{TM}} as the average similarity of this method.
		Best results are highlighted with \colorbox{fg}{light background color}, and the worst \colorbox{bg}{darker background color}.
	}
	\label{crawl:tab:result}
	\centering
	\setlength{\tabcolsep}{4pt}
	\begin{tabular}{@{} c r S[table-format=3.0] S[table-format=5.0] S[table-format=4.0] S[table-format=2.0] S[table-format=2.2] S[table-format=1.2] S[table-format=4.0] S[table-format=3.0] c S[table-format=1.2] @{}}
		\toprule
		 & Method & {S}     & {\abs{P}}  & {\abs{P+}} & {\abs{Seed}} & {HR}                    & {\mymax{\sim}}          & {\abs{Dom}} & {\abs{Dom+}} & TM & {$\avg{\text{TM}}$}      \\
		\midrule
		\multirow{7}{*}{\rotatebox[origin=C]{90}{\num{500}}}
		 & \tcbfk & 128     & 6199       & 387        & 24           & 6.24294241006614        & 0.8996259570121765      & 751         & 96           & F  & \hln 0.10851906973225109 \\
		 & \tcbf  & \hln 11 & 405        & 32         & 6            & 7.901234567901234       & 0.9036480188369751      & 145         & 24           & B  & 0.1749150982023748       \\
		 & \tcfk  &         & \hlp 9110  & 1877       & \hlp 30      & 20.603732162458838      & \hlp 1.000000238418579  & 384         & 23           & F  & 0.24594318970017948      \\
		 & \tcbk  & 278     & 2344       & 339        & 9            & 14.462457337883958      & 0.9036480188369751      & \hlp 873    & \hlp 132     & B  & 0.19825698758933502      \\
		 & \tcf   &         & 9982       & \hlp 2382  & 28           & \hlp 23.862953315968743 & 0.8662863969802856      & 244         & 38           & F  & \hlp 0.28710287588134936 \\
		 & \tcb   & 111     & 1313       & 97         & 12           & 7.387661843107388       & 0.9036480188369751      & 397         & 61           & B  & 0.15058423806970686      \\
		 & \tck   & 21      & \hln 148   & \hln 4     & \hln 1       & \hln 2.702702702702703  & \hln 0.8056498765945435 & \hln 105    & \hln 4       & K  & 0.2357944953077622       \\
		\midrule
		\multirow{7}{*}{\rotatebox[origin=C]{90}{\num{2000}}}
		 & \tcbfk &         & 21663      & 4889       & 46           & 22.568434658172923      & 0.921763002872467       & 1484        & 189          & F  & \hlp 0.37767799866294605 \\
		 & \tcbf  &         & \hlp 26715 & 4965       & \hlp 56      & 18.58506457046603       & 0.921763002872467       & 1683        & \hlp 270     & F  & 0.29014989957614323      \\
		 & \tcfk  &         & 18270      & 4119       & 39           & 22.545155993431856      & \hlp 1.000000238418579  & 809         & 49           & F  & 0.2732494328174037       \\
		 & \tcbk  &         & 8175       & 2055       & 19           & \hlp 25.137614678899084 & 0.9036480188369751      & \hlp 1965   & 269          & B  & 0.2690592301937659       \\
		 & \tcf   &         & 21710      & \hlp 4992  & 48           & 22.994011976047904      & \hlp 1.0000001192092896 & 372         & 52           & F  & 0.2666160293233934       \\
		 & \tcb   & 117     & 1290       & 102        & 12           & 7.906976744186046       & 0.9036480188369751      & 387         & 61           & B  & \hln 0.15412890102541643 \\
		 & \tck   & \hln 22 & \hln 155   & \hln 5     & \hln 1       & \hln 3.225806451612903  & \hln 0.8056498765945435 & \hln 111    & \hln 5       & K  & 0.23670219937744347      \\
		\bottomrule
	\end{tabular}
\end{table*}

\cref{crawl:tab:result} presents the results of the evaluation.
For \num{500} steps we reached a maximum harvest rate of $\sim 23.86\%$ with \tcf, followed by \tcfk{} with $\sim 20.6\%$.
The worst result achieved \tck{} with a harvest rate of just $2.7\%$, which, overall, performed the worst of all combinations.
When backlink search was present in the crawl, it performed the best, expect for \tcbfk.
In terms of domains, \tcbk{} was able to spread the search the widest with \num{873} searched domains with \num{132} of them being relevant, despite crawling only \num{2344} pages~(including the seed).
Those included multiple security overview and dataset pages\footnote{E.g, \url{https://malpedia.caad.fkie.fraunhofer.de} and \url{https://github.com/CyberMonitor/APT_CyberCriminal_Campagin_Collections}.}, as well as multiple unknown security news pages.

During the \num{2000} step test, we saw much more balanced harvest rates over the used methods~(near to or over $20\%$, except \tcb{} and \tck) with forward search gathering the most relevant pages on average.
The top harvest rate was achieved by \tcbk{} with $25.14\%$.
Keyword search alone still performed the worst but combining it with backlink search achieved the highest harvest rate and gathered the highest amount of relevant domains.
In total, nearly a quarter of all crawled pages are of different domains~(with \tcbk) showing the impact of the keyword search through search engines, and we identified in total \num{270} relevant domains~(with \tcbf).

\section{Discussion, Limitations \& Future Work}
\label{crawl:sec:discussion}

\threatcrawl{} effectively addresses the challenge of identifying and crawling \ac{cti} related sources from the web, which answers our research question \say{\crawlermainrq}~\rq{}.
By utilizing a \ac{mab} approach and seed URLs, the system efficiently expands relevant web content, offering a targeted crawling strategy that is well-suited for \ac{cert} and \ac{soc} personnel.
Unlike related crawlers~\cite{pham_bootstrapping_2019,koloveas_intime_2021}, \threatcrawl{} focuses on expanding a predefined set of pages, aligning with the current needs and priorities instead of keywords.
In terms of efficiency, the system significantly outperforms related work.
While \cite{koloveas_intime_2021} reported a harvest rate of $\sim 9.5\%$, our system achieves a much higher rate of~$\sim 25.14\%$, indicating a more effective method for discovering relevant information.
\threatcrawl{}~\c{1} is able to identify key information sources relevant to the initial search domain and even expand the current seed by over 300\% without relaxing its focus.
It went over the course of nearly \num{2000} different web domains and identified \num{270} relevant ones with starting just $17$ seeds~\c{2}.

\paragraph{Limitations}
Despite these positive results, \threatcrawl{} faces some key limitations.
Firstly, it should be noted that the related page search functionality, as described in \cite{zhang_dsdd_2021}, is no longer available due to the removal of this feature by search engines\footnote{Query a search engine with \emph{\say{related:page\_url}} returned related pages, excluding those of the given domain.}.
This removal has resulted in the loss of a key capability for searching across a wide range of sources.
Second, no comparisons with others~\cite{koloveas_intime_2021} was conducted, preventing a broader benchmarking of performance besides the one stated in the references directly.
Runtimes exceeding \num{2000} steps were not evaluated, so we did not observe how the crawler behaved over time or whether it reaches saturation with certain search actions.
Tipping points for the crawling parameters are not evaluated either, \eg, using to lax thresholds, using to distinct seeds, or using to few seeds.
Finally, embeddings are generated using pre-trained \ac{sbert} models rather than larger models like LLaMA or GPT, which could offer improved semantic accuracy but at the cost of higher computational demands and additional privacy considerations.

\paragraph{Future Work}

Future work could include dynamic adjustment of thresholds based on real-time crawler performance, allowing better adaptability.
User feedback from \ac{cert} and \ac{soc} personnel could also be integrated to guide the system’s relevance assessments.
Building on that, while using a \ac{mab} is definitely a top choice to decide, which actions are more promising in the long run and if there is no information of the page that is used for the current search action.
But with crawling this information is just a \texttt{GET} away.
For example, a very detailed page with very few links could yield better results with keyword or backlink search.
Otherwise, if the page is very new, backlink would probably yield worse results than the others.
Using graph-based approaches to map and analyze relationships between crawled pages could provide deeper insights into the structure of the \ac{cti} landscape.
This could be combined with local-sensitive hashing to identify news aggregation platforms and pages that copy others, as well as, the first publisher of information.
Other improvements could be a multistep-depth crawling, \ie, if an irrelevant page is reached keep crawling for $n$ steps just to be sure.
Such an approach could be combined with URL classification~\cite{mahdaouy_domurls_bert_2024} or adaptable domain blacklisting.
While the focus of this paper is the \ac{cti} domain, the system should perform well on other domains too, since it is primarily based on the used seed.
This aspect needs to be evaluated further.

\section{Conclusion}
\label{crawl:sec:conclusion}

\ac{cti} information is published in unstructured form on the web, which presents a time-consuming task for \acp{cert} and \acp{soc} to maintain an up-to-date list of web pages to visit for such information.
Our proposed \threatcrawl{} system addresses the challenge of identifying previously unknown and relevant \ac{cti} sources from the vast amount of unstructured public information available online.
By using a \ac{mab} approach, it efficiently expands a seed of URLs, making it highly suitable for security personnel who need to automate this time-consuming task, even with a small amount of seed URLs $\leq 20$.
With a harvest rate of over 25\%, \threatcrawl{} outperforms prior work, uncovering previously unknown information sources and datasets.
However, limitations like short evaluation runtimes and reliance on \ac{sbert} leave opportunities for further enhancement.
Future work should focus on optimizing search actions, adjusting thresholds dynamically, and leveraging larger models for better accuracy and adaptability.

\ifanonymous\else
	\subsubsection*{\ackname}
This work has been co-funded by the LOEWE initiative (Hesse, Germany) within the emergenCITY centre\footnote{Funding code: [LOEWE/1/12/519/03/05.001(0016)/72]}
and
by the German Federal Ministry of Education and Research and the Hessian Min\-istry of Higher Education, Research, Science and the Arts in their joint support of the
National Research Center for Applied Cy\-ber\-se\-cu\-ri\-ty ATHENE.

\fi

\bibliographystyle{splncs04}
\bibliography{crawler}

\end{document}